\documentclass[twocolumn, showpacs, superscriptaddress, a4paper]{revtex4}%
\usepackage{amsfonts}
\usepackage{amsmath}
\usepackage{amssymb}
\usepackage{dcolumn}
\usepackage{bm}
\usepackage{graphicx}
\usepackage{geometry}%
\providecommand{\U}[1]{\protect\rule{.1in}{.1in}}
\begin{document}
\title{Quantum walks accompanied by spin-flipping in one-dimensional optical lattices}
\author{Li Wang}
\email{liwangiphy@sxu.edu.cn}
\affiliation{Institute of Theoretical Physics, Shanxi University, Taiyuan 030006, P. R. China}
\affiliation{Department of Theoretical Physics, Research School of Physics and Engineering, Australian National University, Canberra, ACT 0200, Australia}
\author{Na Liu}
\affiliation{Institute of Theoretical Physics, Shanxi University, Taiyuan 030006, P. R. China}
\author{Shu Chen}
\affiliation{Beijing National
Laboratory for Condensed Matter Physics, Institute of Physics,
Chinese Academy of Sciences, Beijing 100190, China}
\affiliation{Collaborative Innovation Center of Quantum Matter, Beijing, China}
\author{Yunbo Zhang}
\affiliation{Institute of Theoretical Physics, Shanxi University, Taiyuan 030006, P. R. China}

\date{\today}

\begin{abstract}

We investigate continuous-time quantum walks of two
fermionic atoms loaded  in one-dimensional optical
lattices with on-site interaction and
subjected to a Zeeman field. The quantum walks are accompanied by spin-flipping processes.
We calculate the  time-dependent density
distributions of the two fermions with opposite spins which are initially
positioned on the center site by means of exact numerical
method. Besides the usual fast linear expansion behavior, we find an
interesting spin-flipping induced localization in the time-evolution of  density distributions. We show 
that the fast linear expansion behavior could be restored by simply ramping on the Zeeman field 
or further increasing the spin-flipping strength. The intrinsic origin of this exotic phenomenon is 
attributed to the emergence of a flat band in the single particle spectrum of the system.  
Furthermore, we investigate the effect of on-site interaction on the dynamics of the 
quantum walkers. 
The two-particle correlations are calculated and signal of localization is also shown therein.
A simple potential experimental application of this interesting phenomenon is proposed.

\end{abstract}


% 37.10.Jk Atoms in optical lattices
% 05.60.Gg Quantum transport
% 05.40.Fb Random walks and Levy flights.

\maketitle

\bigskip

\section{INTRODUCTION}

Quantum walk, which is a natural and straightforward generalization of the classical random walk to the quantum world 
\cite{Aharonov,kempe}, has nowadays been a topic of great interest to both theorists 
and experimentalists. A very recent experiment implemented a continuous-time quantum walk 
of two identical bosons on an one-dimensional optical lattice to probe the complex quantum 
phenomena therein \cite{greiner15}. Their approach could be scaled to larger systems,
greatly extending the class of problems accessible via quantum walks.  
Compared to classical random walks, one of those crucial features that quantum walks possessed is the 
quantum coherence, which may lead to much faster expansion of walkers. This advantage of quantum 
walks could be harnessed in fast search algorithm designing for quantum computation \cite{kempe, Childs2009, Childs2013, gutmann}. 
Recently, quantum walk has been found to be very useful in detecting and understanding the topological states
\cite{Kitagawa10,Kitagawa12,Lang,gongjb, Kraus, schen15} and bound states \cite{Fukuhara}. And the interplay 
between interaction  and quantum statistics has also been addressed in the context of two-particle quantum walks 
\cite{qinxz, liwang,Lahini}.

Systems composed of ultracold atoms have been a versatile toolbox for simulating intriguing condensed matter physics for years. Thanks to their high controllability and flexibility, recent breakthrough of generating synthetic gauge fields in ultrcold atomic gases implementing laser techniques has further enabled physicists to touch on spin-related issues on this ultracold atom platform, for example, spin-orbit coupling \cite{spielman,zhangj,cheuk, williams, fuz, madison, ketterle, dalibard,jaksch03, dalibard10, lewenstein14, lewenstein12, spielman09, miyake, bloch, liwang13, goldman,struck}, which plays crucial roles in many important condensed-matter phenomena \cite{kane,zhangsc}.
In this article, we investigate the dynamics of continuous-time quantum walks of two fermionic atoms accompanied by spin-flippings in the hope of shedding some light on the peculiar effects of spin-flipping operations.
Putting aside the absence of time-reversal symmetry and within the paradigm of current cold atom physics, the spin-flipping term we consider here is quite similar to the above mentioned spin-orbit coupling for lattice models \cite{jaksch03, dalibard10, lewenstein14, lewenstein12, spielman09, miyake, bloch,liwang13, goldman}, which thereby may be realized within the same experimental settings, say, laser-assisted tunneling technique \cite{lewenstein14, lewenstein12}.
We vary the strength of the spin-flipping operation and calculate the time-evolving density distribution of the two fermionic atoms among the lattice sites. 
The observation of this simple quantity leads to an interesting localization phenomenon, which is quite anti-intuitive in that one generally expects the spin-flipping operations (could be taken as some kind of tunnelings because of the similarity between their forms) should dramatically enhance the expansion speed of the quantum walkers.  And this is really the case, if we further increase the strength of the spin-flipping  operations  beyond the localization point. The intrinsic origin underlying this exotic phenomenon is addressed basing on an analytical discussion. Inspired by this exotic phenomenon, we propose a new  application of quantum walks in ultracold atom systems. At last, we study the effect of on-site interaction on the dynamics of quantum walkers and calculate the two-particle correlations.

The paper is organized as follows. In Sec. II, we introduce the model featuring spin-flipping processes. On-site interaction and a Zeeman field are also considered. We construct the
Hilbert space for quantum walkers and briefly describe the method we use. The time-evolution of the density distributions under different strengths of spin-flipping operations and Zeeman field is shown in Sec. III and detailed  analysis corresponding to the dynamical properties mentioned is addressed therein.
In Sec. IV, we turn to investigate the dynamics of quantum walks with strong on-site interactions. Two-particle correlations between the two quantum walkers are shown in Sec. V. Finally, a brief summary is given in Sec. VI.

\section{Model and Method\label{secii}}

\begin{figure}[t]
\includegraphics[width=0.35\textwidth]{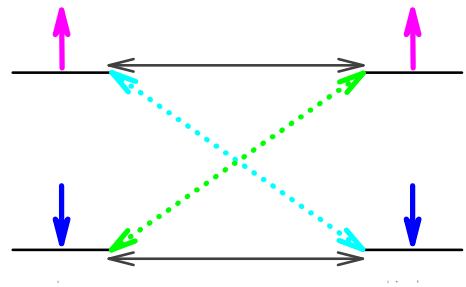} \caption{(Color online)
Sketch of tunnelings  with (dotted) and without (solid) spin-flipping of fermionic atoms between two nearest-neighbouring sites $i$ and $i+1$ in an one-dimensional optical lattice.}%
\label{fig0}%
\end{figure}

We investigate the continuous-time spin-flipping accompanied quantum walks of two fermionic atoms with different spins which are initially prepared on the center site of an one-dimensional optical lattice.  The 1D optical lattice is subjected to an out-of-plane Zeeman field and the on-site interaction between ferimionic atoms with opposite spins is considered. The dynamics of such a system is governed by the standard tight-binding Hamiltonian
\begin{align}
H=H_{t}+H_{sf}+H_{else}.
\label{H}%
\end{align}
The first term describes the hopping of the two-component fermionic atoms in the optical lattice, which reads
\begin{align}
H_{t}=-\sum_{i,\sigma} \left( t_{\sigma\sigma}\hat{c}_{i\sigma}^{\dagger} \hat{c}_{i+1\sigma}+H.c.\right),
\end{align}
where $ t_{\sigma\sigma}$ is the hopping amplitude of spin-$\sigma$ component, $\hat{c}_{i\sigma}$ and $\hat{c}_{i\sigma}^{\dagger}$ are  fermionic annihilation and creation operators, respectively and $\sigma=\uparrow $ or $ \downarrow$ can be regarded as index of pseudospin, which may correspond to one of the hyperfine states of fermonic atoms in cold atom experiments.  The second term $H_{sf}$ corresponds to the spin-flipping operations described by
\begin{align}
H_{sf}=-\sum_{i}\left( \alpha_{\uparrow\downarrow}\hat{c}_{i\uparrow}^{\dagger} \hat{c}_{i+1\downarrow}+
\alpha_{\downarrow\uparrow}\hat{c}_{i\downarrow}^{\dagger} \hat{c}_{i+1\uparrow}+H.c.\right),
\end{align}
where $\alpha_{\uparrow\downarrow}$ and $\alpha_{\downarrow\uparrow}$ are the strengths of spin-flipping operation between  fermionic atoms with opposite spins on the nearest-neighboring sites along diagonal and back-diagonal directions, as depicted in Fig. \ref{fig0}.
The third term reads,
\begin{align}
H_{else}=U\sum_{i}\hat{n}_{i\uparrow} \hat{n}_{i\downarrow} 
-h\sum_{i}\left(\hat{n}_{i\uparrow}-\hat{n}_{i\downarrow}\right),
\end{align}
where $U$ is the on-site interaction,   
$h$ is the out-of-plane Zeeman field, 
and $\hat{n}_{i}=\hat{n}_{i\uparrow}+\hat{n}_{i\downarrow}$ denotes the total particle number, with $\hat{n}_{i\sigma}=\hat{c}_{i\sigma}^{\dagger} \hat{c}_{i\sigma}$ being the paticle number of spin-$\sigma$ fermionic atoms on site $i$.

Straightforwardly, the first two terms $H_{t}$ and $H_{sf}$ can be rewitten in a more compact form
\begin{align}
H_{0}=-\sum_{i,\sigma\sigma'} \left( \hat{c}_{i\sigma}^{\dagger} T_{\sigma\sigma'}\hat{c}_{i+1\sigma'}+H.c. \right),
\end{align}
with the tunneling matrix $T$ being
\begin{align}
T=\left(\begin{array}{cc}  t_{\uparrow\uparrow} & \alpha_{\uparrow\downarrow} \\  \alpha_{\downarrow\uparrow}  & t_{\downarrow\downarrow}\end{array}\right),
\end{align}
which can be designed in practically any form with the aid of laser-assisted tunneling technique \cite{lewenstein12, dalibard10, lewenstein14, jaksch03}.  In experiments, the hopping amplitude,  spin-flipping (spin-orbit coupling, if time-reversal symmetry adopted), Zeeman field, and on-site interaction, may be tuned  \emph{independently}.

Hereafter, we set $t_{\uparrow\uparrow}=t_{\downarrow\downarrow}=1$ as our energy scale throughout  this work. All other parameters are scaled by $t_{\uparrow\uparrow}$$(t_{\downarrow\downarrow})$ in the following numerical and analytical investigations. We consider the initial condition being that two fermionic atoms with opposite spins (or two different hyperfine states) are prepared on the center site of the lattice and the dynamics
on $2L+1$ lattice sites represents a typical quantum walk problem of
continuous-time. We resort to exact numerical techniques to investigate the quantum dynamics of the two-particle quantum walks on 1D spin-flipping accompanied optical lattices.

We now discuss the Hilbert space involved by the quantum walks of two particles. Since
$\left[  N,H\right]  =0$, the total particle number $N=\sum_{i}\hat{n}_i$ is conserved and the
system will evolve in the two-particle Hilbert space. However, the total particle number $N_{\sigma}$ of each spin is no longer conserved due to the existence  of spin-flipping processes. Therefore, the Hilbert space of this system is comprised of three cases, two spin-ups, two spin-downs, and one spin-up with the other spin-down. The spin-flipping process expands the Hilbert space of the original Fermi Hubbard model by coupling the three above mentioned subspaces together. Correspondingly, the Hilbert space is spanned by the basis $B^{(2)}$ $=\{
\left| i\sigma j \sigma' \right>=\left(1- \delta_{ij}\delta_{\sigma\downarrow}\right)\hat{c}_{i\sigma}^{\dagger}  \hat{c}_{j\sigma'}^{\dagger} \left|\mathbf{0}\right> , -L \leqslant i \leqslant j \leqslant  L   \}$. 
Here, $\left\vert \mathbf{0}\right\rangle $ denotes the vacuum state.
With the above basis $B^{(2)}$, it is easy to construct the
Hamiltonian matrix $H$ in two-particle sector. In units of $\hbar=1$,
the time evolution of an arbitrary state obeys%
\begin{equation}
i\frac{d}{dt}\left\vert \psi\left(  t\right)  \right\rangle =H^{\left(
2\right)  }\left\vert \psi\left(  t\right)  \right\rangle , \label{hs}%
\end{equation}
with $\left\vert \psi\left(  t\right)  \right\rangle =\sum_{\sigma\sigma',i\leq j%
}C_{i\sigma j\sigma'}\left(  t\right)  \left\vert i\sigma j\sigma' \right\rangle $ for
two-component fermions. Below we will study the continuous-time quantum walks of two fermions
starting from an initial state $\left\vert
\psi_{ini}\right\rangle =\hat{c}_{0\uparrow}^{\dagger}\hat{c}_{0\downarrow}^{\dagger}\left\vert
\mathbf{0}\right\rangle $.
In order to explore the dynamical behavior of the quantum walkers, we calculate the time-dependent density distribution $\left < n_{i\sigma}(t) \right>$ of fermionic atoms with spin $\sigma$ in the whole lattice sites at time $t$
\begin{align}
\left < n_{i\sigma}(t) \right> = \left< \psi (t) \left | \hat{c}_{i\sigma}^{\dagger}  \hat{c}_{i\sigma} \right |  \psi (t) \right>,
\end{align}
and that of both spin-up and spin-down fermionic atoms
\begin{align}
\left<  n_{i}(t) \right>=\left<  n_{i\uparrow}(t) \right>+\left<  n_{i\downarrow}(t) \right>.
\end{align}
Besides, in order to have a look at the effects of quantum statistics, we calculate the two-particle correlation
\begin{align}
\Gamma_{ij}(t) = \left< \psi (t) \left | \hat{c}_{i\uparrow}^{\dagger} \hat{c}_{j\downarrow}^{\dagger}  \hat{c}_{j\downarrow}\hat{c}_{i\uparrow}  \right |  \psi (t) \right>
\end{align}
at the end of the quantum walks.
Corresponding results are shown in the Sec. \ref{tpc}. Also, two-particle correlations of the system's eigenstates are calculated therein. 

\begin{figure}[t]
\includegraphics[width=0.5\textwidth]{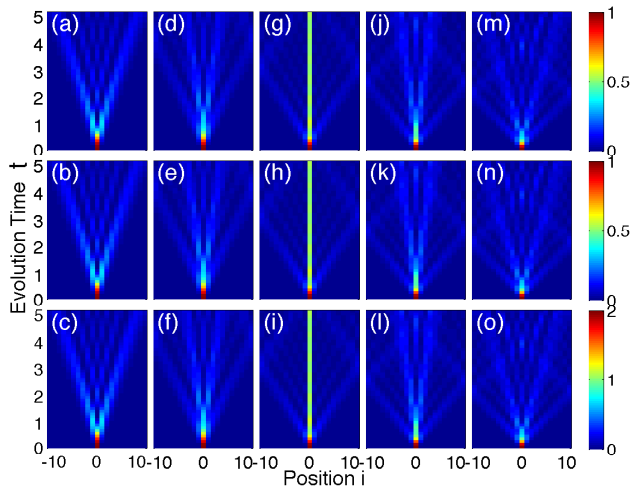} \caption{(Color online)
Quantum walks of two fermionic atoms with different spins. The first row is
the density distribution of atoms with spin-up, the second row is that of atoms with spin-down,
and the third row is the total density distribution of both spin-up and spin-down.
For all these figures, we have $t_{\uparrow\uparrow}=t_{\downarrow\downarrow}=1$,
and $U=h=0$.
(a-c) $\alpha_{\uparrow\downarrow}=\alpha_{\downarrow\uparrow}=0.0$,
(d-f) $\alpha_{\uparrow\downarrow}=\alpha_{\downarrow\uparrow}=0.5$,
(g-i) $\alpha_{\uparrow\downarrow}=\alpha_{\downarrow\uparrow}=1.0$,
(j-l) $\alpha_{\uparrow\downarrow}=\alpha_{\downarrow\uparrow}=1.5$,
(m-o) $\alpha_{\uparrow\downarrow}=\alpha_{\downarrow\uparrow}=2.0$.}%
\label{fig1}
\end{figure}

\section{spin-flipping accompanied quantum walks on 1D optical lattices} \label{sfqw}

We first investigate the quantum walks of two fermionic atoms without interaction and in absence of Zeeman field, i.e. $U=h=0$. As shown in Fig. \ref{fig1}(a-c),  the well-known linear time-dependent expansion behavior appears when there is no spin-flipping operations in the system. When we turn on the spin-flipping process with a strength of $\alpha_{\uparrow\downarrow}=\alpha_{\downarrow\uparrow}=0.5$, the time-evolution of the density distribution splits into two branches (Fig.\ref{fig1}(d-f)). Both of them obey the linear expansion behavior against time $t$, however, one is faster, the other is slower. Then we further increase the strength of spin-flipping process to $\alpha_{\uparrow\downarrow}=\alpha_{\downarrow\uparrow}=1.0$ (see Fig. \ref{fig1}(g-i)), an interesting localization on the initial site comes into view with a much faster linear expansion as a background.
Subsequently, we further increase the strength of the spin-flipping term and find that the exotic localization in the particle density distribution shrinks gradually, see Fig. \ref{fig1}(j-l). The time evolution of the density distribution displays two branches as that of Fig. \ref{fig1}(d-f), however the outer branch is becoming much faster. Even stronger spin-flipping term will further speed up the expansion of  the quantum walkers, as shown in Fig. \ref{fig1}(m-o).

\begin{figure}[t]
\includegraphics[width=0.5\textwidth]{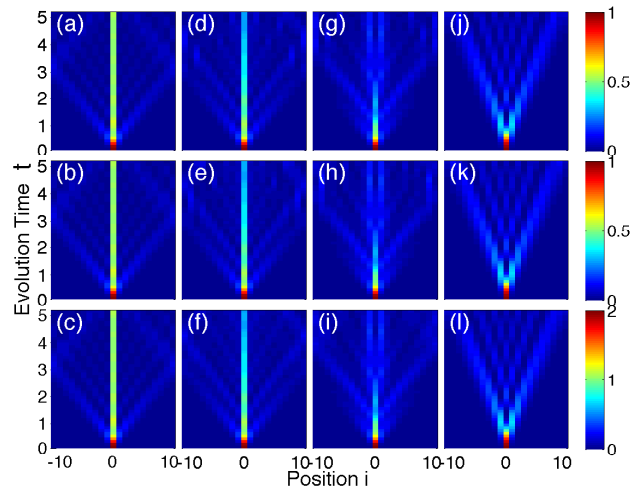} \caption{(Color online)
The restoration of linear expansion behavior in quantum walks by ramping on
Zeeman field. The first row is
the density distribution of atoms with spin-up, the second row is that of atoms with spin-down,
and the third row is the total density distribution of both spin-up and spin-down.
For all these figures, we have $t_{\uparrow\uparrow}=t_{\downarrow\downarrow}=1$, $\alpha_{\uparrow\downarrow}=\alpha_{\downarrow\uparrow}=1.0$,
and $U=0$.
(a-c) $h=0.0$,
(d-f) $h=0.5$,
(g-i) $h=1.0$,
(j-l) $h=10.0$.}
\label{fig2}
\end{figure}

 Also, we investigate the effect of the outer-of-plane Zeeman field. It is found that the interesting localization could be destroyed by Zeeman field and the expected linear expansion behavior restores. According to Fig. \ref{fig2}, we can see that, as the Zeeman field grows stronger, the localized part grows weaker and weaker and finally the linear expansion behavior reappears. However, the expansion of the quantum walkers is much slower than the previous background expansion, i.e. the Zeeman field could slow down the propagation speed of quantum walkers. For all the pictures in Fig. \ref{fig2}, we have $t_{\uparrow\uparrow}=t_{\downarrow\downarrow}=\alpha_{\uparrow\downarrow}=\alpha_{\downarrow\uparrow}=1.0$ and
$U=0$.

The above-mentioned exotic localization stimulates one's desire to find out the origin underlying this interesting phenomenon. After some tries,
we finally find out that this exotic spin-flipping induced localization is intrinsically  related to the structure of the system's energy spectrum. For the system without interactions and under periodic boundary condition (PBC), its energy spectrum is easily calculated through the discrete Fourier transformation
\begin{align}
\hat{c}_{n\sigma}^{\dagger}=\frac{1}{\sqrt{2L+1}} \sum_{k} e^{ikn}\hat{c}_{k\sigma}^{\dagger},
\end{align}
where the quasimomentum $k$ is constrained to the first Brillouin Zone and $n$ is the lattice site index. Straightforwardly,
the Hamiltonian Eq. (\ref{H}) without interaction could be written in momentum space as
\begin{align}
H=\sum_{k} \hat{\Psi}_{k}^{\dagger}H_{k} \hat{\Psi}_{k},
\end{align}
in which $\hat{\Psi}_{k}^{\dagger}=( \hat{c}_{k\uparrow}^{\dagger},\hat{c}_{k\downarrow}^{\dagger} ) $, and
\begin{align}
H_k=-\left(\begin{array}{cc} t_{\uparrow\uparrow}e^{-ik}+t_{\uparrow\uparrow}^{*}e^{ik}+h &
\alpha_{\uparrow\downarrow} e^{-ik}+\alpha_{\downarrow\uparrow}^{*} e^{ik} \\
\alpha_{\downarrow\uparrow} e^{-ik}+\alpha_{\uparrow\downarrow}^{*} e^{ik}
 & t_{\downarrow\downarrow}e^{-ik}+t_{\downarrow\downarrow}^{*}e^{ik}-h \end{array}\right).
 \label{Hk}
\end{align}
Within the Nambu scheme,  energy spectrum is obtained simply by diagonalizing $H_k$ by hand.

\begin{figure}[t]
\includegraphics[width=0.48\textwidth]{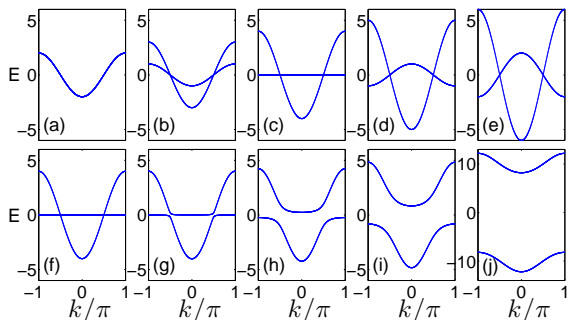} \caption{
Single-particle spectra of the system in momentum space.
(a-e) Effects of spin-flipping operations and the emergence of a flat band. 
(a) $\alpha_{\uparrow\downarrow}=\alpha_{\downarrow\uparrow}=0$, 
(b) $\alpha_{\uparrow\downarrow}=\alpha_{\downarrow\uparrow}=0.5$,
(c) $\alpha_{\uparrow\downarrow}=\alpha_{\downarrow\uparrow}=1$,
(d) $\alpha_{\uparrow\downarrow}=\alpha_{\downarrow\uparrow}=1.5$,
(e) $\alpha_{\uparrow\downarrow}=\alpha_{\downarrow\uparrow}=2$.
(f-j) melting of the flat band by Zeeman field. 
(f) $h=0$, (g) $h=0.2$, (h) $h=1$, (i) $h=2$, (j) $h=10$.
For all pictures above, $ t_{\uparrow\uparrow}=t_{\downarrow\downarrow}=1$.}
\label{fig3}
\end{figure}

As shown in Fig. \ref{fig3}(a),  in absence of spin-flipping operations, the energy spectrum of the spin-up component and  that of the spin-down component are independent of each other and coincide.  When the spin-flipping process is introduced in, the spectra of the two component fermions no longer coincide, instead, one is becoming wider and the other is crushed narrower in the dimension of energy, see Fig. \ref{fig3}(b). This, in fact, is corresponding to the two branches of the density distribution evolution in Fig. \ref{fig1}(d-f).  As the strength of spin-flipping along diagonal and back-diagonal directions shown in Fig. \ref{fig0} becomes of the same strength of tunnelings, i.e., $\alpha_{\uparrow\downarrow}=\alpha_{\downarrow\uparrow}=t_{\uparrow\uparrow}=t_{\downarrow\downarrow}=1$, a flat-band emerges in the spectrum, see Fig. \ref{fig3}(c). This is close related to the appearance of the exotic spin-flipping induced localization mentioned above.
However, if we go on increasing the strength of the spin-flipping operations, the flat band gradually expands to the opposite direction, and transforms into a normal band, See Fig. \ref{fig3}(d,e). In the whole process in Fig. \ref{fig3} (a-e), the other band is expanding wider and wider all the time, which drives the background expansion in Fig. \ref{fig1} faster and faster.

\begin{figure}[t]
\includegraphics[width=0.48\textwidth]{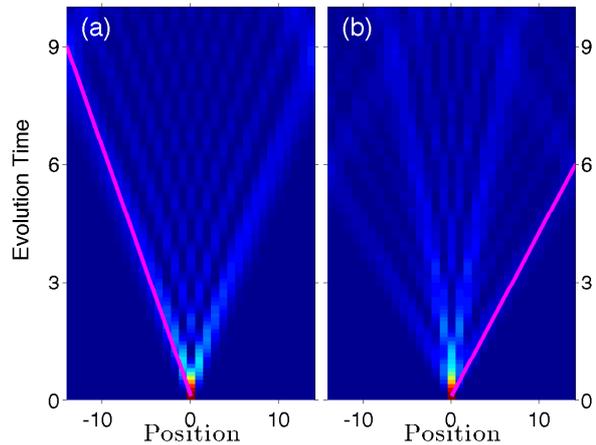} \caption{(Color online)
Quantum walks accompanied by spin-flipping of strength $\alpha_{\uparrow\downarrow}=\alpha_{\downarrow\uparrow}=\alpha=0$(a) and $\alpha_{\uparrow\downarrow}=\alpha_{\downarrow\uparrow}=\alpha=0.5$(b).
The line is used as a guide of eye. The shortest time needed for quantum walkers to reach the lattice boundary is about 9.0(a) and 6.0(b).}
\label{figs04}
\end{figure}

To be more quantitative, here we investigate the dependence of the propagation speed of the quantum walkers on the tunnelling and spin-flipping parameters.
For the case of $t_{\uparrow\uparrow}=t_{\downarrow\downarrow}=t$, $\alpha_{\uparrow\downarrow}=\alpha_{\downarrow\uparrow}=\alpha$ and $U=h=0$,  Eq. (\ref{Hk}) is simplified as 
\begin{align}
H_k=-2\cos{k}\left(\begin{array}{cc} t & 
\alpha  \\
\alpha
 & t \end{array}\right).
\end{align}
Straightforwardly, the eigenenergies of the system are obtained as,
\begin{align}
E_{k}=-2 \left(t\pm \alpha \right) \cos{k} .
\end{align}
According to the traditional band theory, the group velocity $v_g$ of particles is proportional to the derivative of $E_k$,
\begin{align}
v_g \propto \frac{\partial{E_k}}{\partial k}=2 \left(t\pm \alpha \right) \sin{k} .
\end{align}
This is just the intrinsic reason for the two speeds in the quantum walks, which is displayed  in the numerical results in Fig. (\ref{fig1}). The propagation velocity of the quantum walkers depends on the value of $t\pm \alpha$.
To make this clearer, a simple example is given. Here
the tunneling parameter is set to $t=1$. Then for quantum walks with spin-flipping operations of strength $\alpha=0$ and $\alpha=0.5$, the ratio between their propagation velocity of the outer branch is proportional to 2/3. Correspondingly, the shortest time needed for them to reach the lattice boundary will be about 3 vs. 2. This is in agreement with the numerical results, as is shown in Fig. \ref{figs04}.

Now we turn to check the effects of Zeeman field on the flat band. As the Zeeman field grows up, a gap opens at $k=\pm \pi/2$ (Fig. \ref{fig3}g). Correspondingly, the flat band is broken into fractions.  And these flat fractions gradually vanish (Fig. \ref{fig3}(h-j)), as the Zeeman field becomes stronger and stronger. Therefore, the restoration of the linear expansion behavior in quantum walks is intrinsically related to the vanishing of the flat band.

According to the discussion above, one could naturally realize that there is a link between the exotic spin-flipping induced localization and the presence of flat band in the system's energy spectra. Therefore, we propose that quantum walks may be used to detect the existence of flat bands, in addition to the detection of bound states\cite{Fukuhara}  and topological states  \cite{Kitagawa10,Kitagawa12,Lang, gongjb, Kraus}. An experimental signal of flat bands is simply the presence of localization in the time-dependent density distribution of quantum walkers.

\section{Effects of interaction on quantum walks}

\begin{figure}[t]
\includegraphics[width=0.5\textwidth]{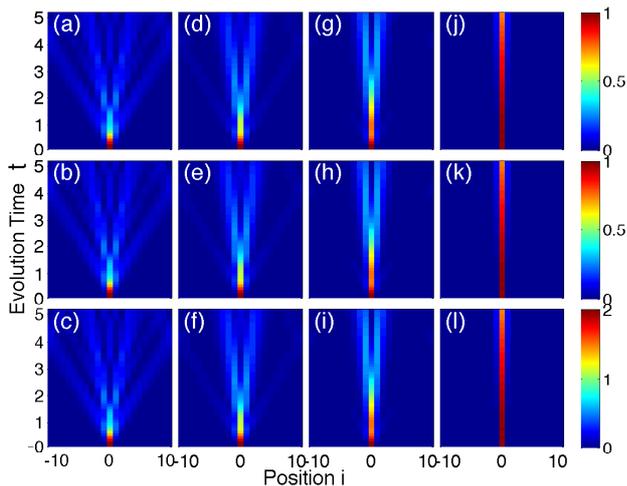} \caption{(Color online)
Localization induced by ramping on the on-site interaction with
spin-flipping of  normal strength. The first row is
the density distribution of atoms with spin-up, the second is that of atoms with spin-down,
and the third row is the total density distribution of both spin-up and spin-down.
For all these figures, we have $t_{\uparrow\uparrow}=t_{\downarrow\downarrow}=1$,
$\alpha_{\uparrow\downarrow}=\alpha_{\downarrow\uparrow}=0.5$,
and $h=0$.
(a-c) $U=0.0$,
(d-f) $U=3.0$,
(g-i) $U=5.0$,
(j-l) $U=20.0$.
}
\label{fig4}
\end{figure}

In this section, we turn to investigate the strongly correlated quantum walks of two fermionic atoms with different spins.  And we focus on the effects of on-site interaction on dynamics of the quantum walkers. The initial state is chosen to be the same as in Sec. \ref{sfqw}.
As shown in Fig. \ref{fig4}, with the increasing of the on-site interaction, the outer branch of the density distribution evolution in Fig. \ref{fig4}(a-c)  becomes fainter and fainter, and finally disappears. This means spreading of the quantum walkers is dramatically suppressed by on-site interaction. On the other hand, the inner branch is being crushed narrower and narrower,  and localization prevails at last.  In essence, this is also a kind of the so-called interaction-induced self-trapping phenomenon\cite{Wang-epjd}.  For all these pictures, we have $t_{\uparrow\uparrow}=t_{\downarrow\downarrow}=1$, $\alpha_{\uparrow\downarrow}=\alpha_{\downarrow\uparrow}=0.5$ and $h=0$.

\begin{figure}[t]
\includegraphics[width=0.5\textwidth]{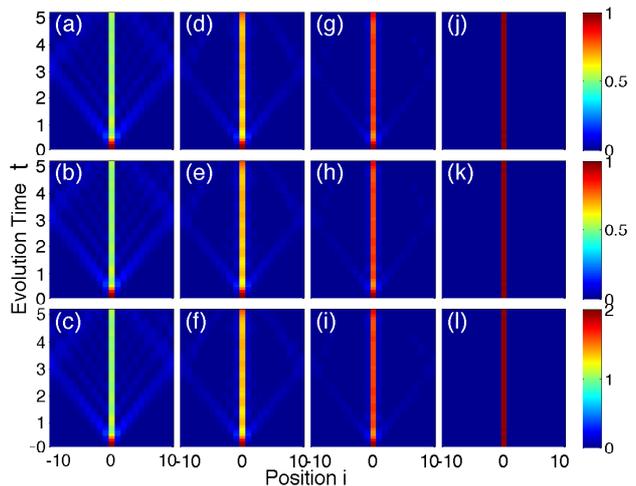} \caption{(Color online)
Intensified localization under the cooperation of strong interaction and flat band effects.
The first row is
the density distribution of atoms with spin-up, the second is that of atoms with spin-down,
and the third row is the total density distribution of both spin-up and spin-down.
For all these figures, we have $t_{\uparrow\uparrow}=t_{\downarrow\downarrow}=1$,
$\alpha_{\uparrow\downarrow}=\alpha_{\downarrow\uparrow}=1.0$,
and $h=0$.
(a-c) $U=0.0$,
(d-f) $U=3.0$,
(g-i) $U=5.0$,
(j-l) $U=20.0$.}
\label{fig5}
\end{figure}

\begin{figure}[b]
\includegraphics[width=0.48\textwidth]{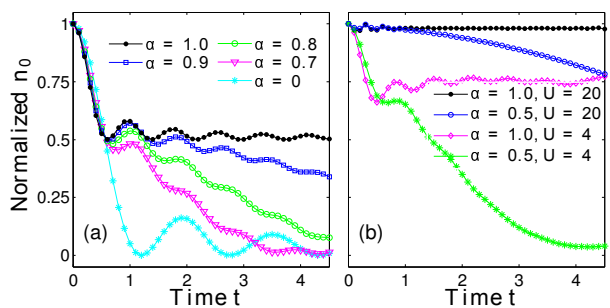} \caption{(Color online)
Time evolution of the localized proportion $n_{0}$ of the two quantum walkers for some set of parameters.
The initial state is chosen as both of the two quantum walkers being positioned on the center site. (a) $U=h=0$. (b) $h=0$. For all these figures, we have $\alpha_{\uparrow\downarrow}=\alpha_{\downarrow\uparrow}=\alpha$.}
\label{figs03}
\end{figure}

In Fig. \ref{fig5}, we show the intensified localization induced by strong on-site interaction with the special spin-flipping strength of $\alpha_{\uparrow\downarrow}=\alpha_{\downarrow\uparrow}=1$. As expected, with the same set of on-site interactions,  the evolution approaching to localization is much faster.   Starting from the spin-flipping induced localization(Fig. \ref{fig5}(a-c)), the original background linear expansion of quantum walkers is going to vanish step by step as the on-site interaction grows up.  And finally, as shown in Fig. \ref{fig5}(j-l) only a localization on the initial site remains, which is dramatically intensified comparing to Fig. \ref{fig4}(j-l).
That is to say,  upon the point $t_{\uparrow\uparrow}=t_{\downarrow\downarrow}=\alpha_{\uparrow\downarrow}=\alpha_{\downarrow\uparrow}=1.0$ where flat band presents, strong on-site interactions dramatically enhance the process towards complete localization.

To see this more clearly,  we calculate the time evolution of the localized proportion $n_{0}$ of the two quantum walkers for some set of parameters, as is shown in Fig. \ref{figs03}. This quantity is normalized by the total particle number.
The results for quantum walks without interaction is also displayed here in Fig. \ref{figs03}(a) for comparison, in which we can see that
as the strength of the spin-flipping operations approaches to the point $\alpha_{\uparrow\downarrow}=\alpha_{\downarrow\uparrow}=\alpha=1.0$, the exotic spin-flipping induced localization gradually appears. Results for strongly interacting quantum walks are shown in Fig. \ref{figs03}(b), comparing the cases with same strength of spin-flipping operations, it is found that the strong on-site interaction can dramatically intensify the spin-flipping induced localization.

\begin{figure}[t]
\includegraphics[width=0.46\textwidth]{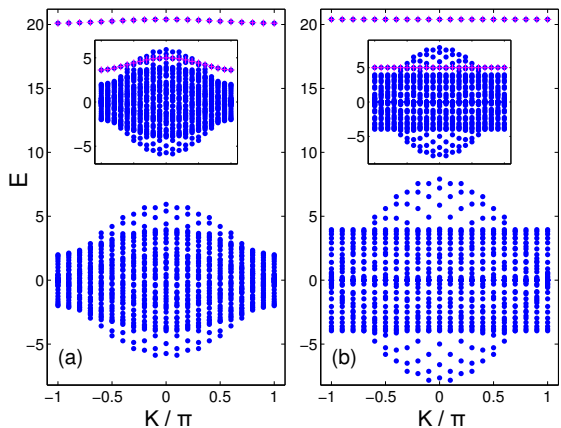} \caption{(Color online)
Two-particle energy spectrum of the system in momentum space with on-site interaction vs total quasi-momentum $K$.
(a) $\alpha_{\uparrow\downarrow}=\alpha_{\downarrow\uparrow}=0.5$, $U=20$ (inset with $U=3$),
(b) $\alpha_{\uparrow\downarrow}=\alpha_{\downarrow\uparrow}=1.0$,  $U=20$ (inset with $U=3$).}
\label{fig6}
\end{figure}

To get a deep understanding, in Fig. \ref{fig6} we show the corresponding two-particle energy spectra in momentum space of finite lattices with $21$ sites. Strong enough repulsive on-site interaction will generate a band of repulsively bound states which is well separated from the continuum band regime \cite{Wang-epjd} in the energy spectrum of the system, see Fig. \ref{fig6}(a) with $U=20$. This underlies the emergence of on-site interaction induced localization in Fig. \ref{fig4}(j-l). A similar phenomenon is the observation of repulsively bound atom pairs
in an optical lattice \cite{Winkler}. While for an on-site interaction of medium strength $U=3$ as shown in the inset of Fig. \ref{fig6}(a), the corresponding band of bound states is mixed with the continuum bands block. For repulsive on-site interaction $U=20$ and spin-flipping strength of $\alpha_{\uparrow\downarrow}=\alpha_{\downarrow\uparrow}=1.0$, a similar highly excited band of bound states is also present in the energy spectrum, see Fig. \ref{fig6}(b). However, in this case, the band of bound states is much flatter comparing to Fig. \ref{fig6}(a). This is reminiscent of the flat band for non-interacting quantum walks discussed in the previous section. Also, series of distinct flat bands is shown in Fig. \ref{fig6}(b) and its inset. We can conclude that presence of flat band and strong on-site interaction together contribute a lot in the process approaching to complete localization.
Moreover, we also investigate strongly interacting quantum walks of the two fermionic atoms with attractive on-site interactions. Similar dynamics of the quantum walkers is found.

\section{Two-particle correlatons \label{tpc}}

\begin{figure}[b]
\includegraphics[width=0.49\textwidth]{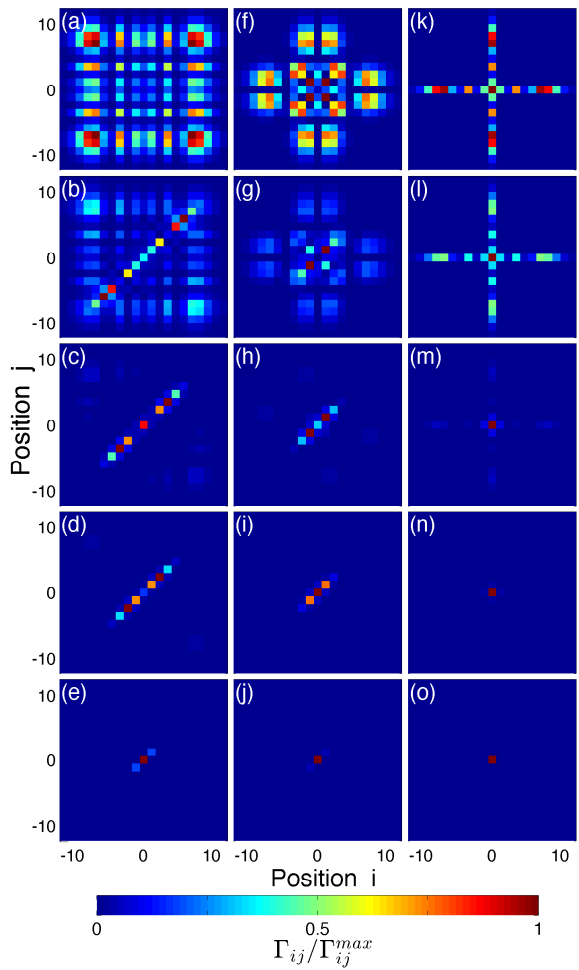} \caption{(Color online)
Two-particle correlations of quantum walkers in 1D optical lattices at the end of the quantum walks. 
From top to bottom, the on-site interaction $U$ are $0$, $1$, $3$, $5$, and $20$, respectively. 
(a-e) $\alpha_{\uparrow\downarrow}=\alpha_{\downarrow\uparrow}=0.0$
(f-j) $\alpha_{\uparrow\downarrow}=\alpha_{\downarrow\uparrow}=0.5$. 
(k-o) $\alpha_{\uparrow\downarrow}=\alpha_{\downarrow\uparrow}=1.0$. For all these figures, we have $h=0$.}
\label{figs01}
\end{figure}

Finally, in this section we calculate the two-particle correlations between the two quantum walkers to have a look at the effects of quantum statistics. Corresponding results are shown in Fig. \ref{figs01}, in which the spin-flipping strength varies among columns and the on-site interaction varies among rows. From left to right,  the values of spin-flipping strength are $0$, $0.5$, $1.0$. From top to bottom, the on-site interaction $U$ are $0$, $1$, $3$, $5$, $20$.
In the first column of Fig. \ref{figs01}, we show the two particle correlations $\Gamma_{ij}$ for the two free ($U=0$) fermionic quantum walkers at the end of the quantum walks before they have reached the lattice boundaries.  In the absence of spin-flipping term, after propagation each particle can be found on either side of the centre site , reflected in the four symmetric peaks in Fig. \ref{figs01}(a). As the strength of the spin-flipping term grows up, the two-particle correlations are compressed. And when the spin-flipping term grows to a strength of $\alpha_{\uparrow\downarrow}=\alpha_{\downarrow\uparrow}=1.0$, the two-particle correlation  transforms into two lines perpendicular to each other and the interesting spin-flipping induced partial localization appears, which is reflected by the central peak surrounded by four weaker peaks in Fig. \ref{figs01}(k). 

Interacting quantum walks of the two fermionic particles under different spin-flipping strengths are shown in the columns of Fig. \ref{figs01}. When we turn on the on-site interaction $U$ between the two quantum walkers and continually increase the strength, the two particles tend to propagate as a pair and finally the density distribution localized, see Fig. \ref{figs01}(a-e), (f-j) and (k-o).  In Fig. \ref{figs01}(k-o), we have  $\alpha_{\uparrow\downarrow}=\alpha_{\downarrow\uparrow}=1.0$ upon which a flat band emerges in the system's spectrum. It is found that the process approaching to localization is intensified by the strong on-site interactions.

\begin{figure}[t]
\includegraphics[width=0.45\textwidth]{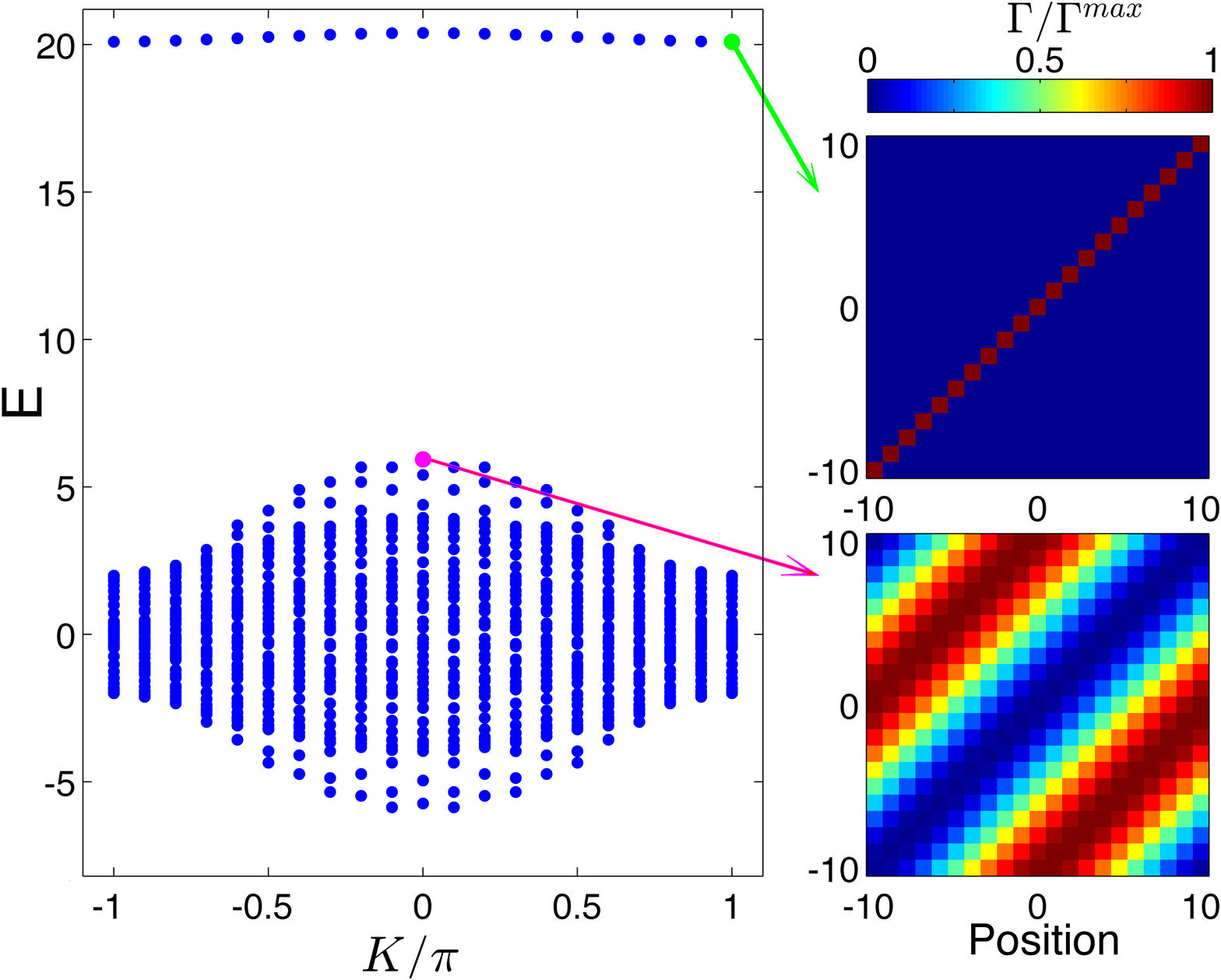} \caption{(Color online)
Energy spectrum of the system and the corresponding two-particle correlations with $\alpha_{\uparrow\downarrow}=\alpha_{\downarrow\uparrow}=0.5$, $U=20$ and $h=0$. (top inset) Two-particle correlation for an eigenstate in the highly excited band. (lower inset) Two-particle correlation for an eigenstate in the lower band. }
\label{figs02}
\end{figure}

In Fig. \ref{figs02}, we calculate the two-particle correlation $\Gamma_{ij}$ for the eigenstates of the system.  It is shown that the highly excited band consists of bound pair states in which two quantum walkers only occupy the same sites. The two-particle correlation matrix only has nonzero values on the diagonal elements. On the other hand, the continuum block in the energy spectrum consists of states in which the two walkers have small probability to occupy the same lattice site, which manifests them as scattering states.

\section{CONCLUSIONS}

In summary, we have investigated the spin-flipping accompanied continuous-time quantum walks of two fermionic atoms with different spins confined in one-dimensional optical lattices and subjected to a Zeeman fields using the exact numerical method. The initial state of the quantum walkers is chosen as both of the two fermionic atoms locating on the center site of the optical lattice. The time-evolution of density distribution of quantum walkers is calculated, in which, the expected linear expansion behavior appears. However, by varying the strengths of the spin-flipping processes,  we are able to observe an exotic spin-flipping induced localization phenomenon. Since the spin-flipping operations considered in this paper may be experimentally realized by using laser-assisted tunneling technique, we expect this interesting localization to be observed directly in the near future experiments. After some tries, we find that the fast linear expansion behavior could be restored by ramping on the Zeeman field to certain strength. Additionally, further increasing the strength of the spin-flipping process can also destroy this interesting localization. In order to understand this exotic localization phenomenon, we resort to discrete Fourier transformation to calculate single-particle spectrum of the system in momentum space. A link between the phenomenon of spin-flipping induced localization and the presence of a flat band in the energy spectrum is found. The observation of such localization in a quantum walk experiment could indicate that a flat band is present. We therefore propose that quantum walks could be taken as a tool to detect flat bands in system's spectrum experimentally. Quantum walks under strong interactions are also addressed and two-particle correlations are calculated. We found that 
strong repulsive interactions enhance localization when the systems has a flat band.

\begin{acknowledgments}
This work is supported by NSF of China under Grant Nos. 11234008, 
11474189 and 11404199, the National Basic Research Program of China (973 Program) under
Grant No. 2011CB921601, Program for Changjiang Scholars and Innovative
Research Team in University (PCSIRT)(No. IRT13076), NSF for youths of Shanxi Province No. 2015021012, and research initiation funds from SXU No. 216533801001. S. C. is supported by NSF of China under Grant Nos. 11425419, 11374354 and 11174360.   The work is partially completed during the visit in ANU which is sponsored by SXU and CSC(No. 201508140015). L. W. would like to thank the hospitality of Profs. V. Bazhanov, A. Truscott and X.-W. Guan during the stay in ANU.
\end{acknowledgments}

%\newpage


\begin{thebibliography}{99}


\bibitem {Aharonov}Y. Aharonov, L. Davidovich, and N. Zagury, Phys. Rev. A
\textbf{48}, 1687 (1993).

\bibitem{kempe} J. Kempe, Comtemporary Physics \textbf{44}, 307 (2003).

\bibitem{greiner15}  P.M. Preiss, R. Ma, M. E. Tai, A. Lukin, M. Rispoli, P. Zupancic, R. Islam, and M. Greiner, Science \textbf{347}, 1229 (2015).


\bibitem {Childs2009}A. M. Childs, Phys. Rev. Lett. \textbf{102}, 180501 (2009).

\bibitem {Childs2013}A. M. Childs, D. Gosset, and Z. Webb, Science \textbf{339},
791 (2013).

\bibitem{gutmann} E. Farhi and S. Gutmann, Phys. Rev. A \textbf{58}, 915 (1998).

\bibitem{Kitagawa10} T. Kitagawa, M. S. Rudner, E. Berg, and E. Demler, Phys. Rev. A \textbf{82}, 033429 (2010).

\bibitem {Kitagawa12}T. Kitagawa, M. A. Broome, A. Fedrizzi, M. S. Rudner, E.
Berg, I. Kassal, A. Aspuru-Guzik, E. Demler, and A. G. White, Nature Comm.
\textbf{3}, 882 (2012).

\bibitem{Lang} L-J. Lang, X. Cai, and S. Chen, Phys. Rev. Lett. {\bf 108}, 220401 (2012).

\bibitem{gongjb} D. Y. H. Ho and J. Gong, Phys. Rev.  Lett. \textbf{109}, 010601 (2012).

\bibitem {Kraus}Y. E. Kraus, Y. Lahini, Z. Ringel, M. Verbin, and O. Zilberberg, Phys. Rev. Lett. \textbf{109}, 106402 (2012).

\bibitem{schen15} L. Li, S. Chen, Euro. Phys. Lett. \textbf{109}, 40006 (2015).

\bibitem{Fukuhara}T. Fukuhara, P. Schau\ss ,\ M. Endres, S. Hild, M.
Cheneau, I. Bloch, and C. Gross, Nature (London) \textbf{502}, 76 (2013).


\bibitem{qinxz} X. Qin, Y. Ke, X. Guan, Z. Li, N. Andrei, and C. Lee, Phys. Rev. A \textbf{90} 062301 (2014).

\bibitem{liwang} L. Wang, L. Wang, and Y. Zhang, Phys. Rev. A \textbf{90}, 063618 (2014).

\bibitem {Lahini}Y. Lahini, M. Verbin, S. D. Huber, Y. Bromberg, R.
Pugatch, and Y. Silberberg, Phys. Rev. A \textbf{86}, 011603 (2012).




\bibitem{spielman} Y.-J. Lin, K. Jim\'enez-Garc\'ia, and I. B. Spielman, Nauture (London) \textbf{471}, 83 (2011).

\bibitem{zhangj} P. Wang, Z.-Q. Yu, Z. Fu, J. Miao, L. Huang, S. Chai, H. Zhai, and J. Zhang, Phys. Rev. Lett. \textbf{109}, 095301 (2012).

\bibitem{cheuk} L. W. Cheuk, A. T. Sommer, Z. Hadzibabic, T. Yefsah, W. S. Bakr, and M. W. Zwierlein, Phys. Rev. Lett. \textbf{109}, 095302 (2012).

\bibitem{williams} R. A. Williams, M. C. Beeler, L. J. LeBlanc, K. Jim\'enez-Garc\'ia, and I. B. Spielman, Phys. Rev. Lett. \textbf{111}, 095301 (2013).

\bibitem{fuz} Z. Fu, L. Huang, Z. Meng, P. Wang, L. Zhang, S. Zhang, H. Zhai, P. Zhang, and J. Zhang, Nat. Phys. \textbf{10}, 110 (2014).



\bibitem{madison} K. W. Madison, F. Chevy, W. Wohlleben, and J. Dalibard, Phys. Rev. Lett. \textbf{84}, 806 (2000).

\bibitem{ketterle} J. R. Abo-Shaeer, C. Raman, J. M. Vogels, and W. Ketterle, Science \textbf{292}, 476 (2001).

\bibitem{dalibard} J. Dalibard, F. Gerbier, G. Juzeli\={u}nas, and P. \"Ohberg, Rev. Mod. Phys. \textbf{83}, 1523 (2011).


%\vspace{12cm}




\bibitem{jaksch03} D. Jaksch, P. Zoller, New J. Phys. \textbf{5}, 56(2003).

\bibitem{dalibard10} F. Gerbier, and J. Dalibard, New J. Phys \textbf{12} 033007 (2010).



\bibitem{lewenstein14} A. Celi, P. Massignan, J. Ruseckas, N. Goldman, I. B. Spielman, G. Juzeli\=unas, and M. Lewenstein, Phys. Rev. Lett \textbf{112}, 043001 (2014).

\bibitem{lewenstein12} L. Mazza, A. Bermudez, N. Goldman, M. Rizzi, M. A. Martin-Delgado and M. Lewenstein, New J. Phys. \textbf{14}, 015007 (2012).

\bibitem{spielman09} Y.-J. Lin, R. L. Compton, K. Jim\'enez-Garc\'ia, J. V. Porto, and I. B. Spielman, Nauture (London) \textbf{462}, 628 (2009).

\bibitem{miyake} H. Miyake, G. A. Siviloglou, C. J. Kennedy, W. C. Burton, and W. Ketterle, Phys. Rev. Lett. \textbf{111}, 185302 (2013).

\bibitem{bloch} M. Aidelsburger, M. Atala, M. Lohse, J.T. Barreiro, B. Paredes, and I. Bloch, Phys. Rev. Lett. \textbf{111}, 185301 (2013).

\bibitem{liwang13} L. Wang, L. Fu, Phys. Rev. A. \textbf{87}, 053612 (2013).

\bibitem{goldman} N. Goldman, A. Kubasiak, A. Bermudez, \emph{et al.}, Phys. Rev. Lett. \textbf{103}, 035301 (2009).

\bibitem{struck}  J. Struck, C. \"Olschl\"ager, R. Le Targat, P. Soltan-Panahi, A. Eckardt, M. Lewenstein, P. Windpassinger, and K. Sengstock, Science \textbf{333}, 996 (2011).



\bibitem{kane} M. Z. Hasan and C. L. Kane, Rev. Mod. Phys. \textbf{82}, 3045 (2010).

\bibitem{zhangsc} X.-L. Qi and S.-C. Zhang, Rev. Mod. Phys. \textbf{83}, 1057 (2011).


\bibitem{jochim15}  S. Murmann, A. Bergschneider, V. M. Klinkhamer, G. Z\"urn, T. Lompe,  and S. Jochim, Phys. Rev. Lett. \textbf{114}, 080402 (2015).


\bibitem{Wang-epjd} L. Wang, Y. Hao and S. Chen, Eur. Phys. J. D \textbf{48}, 229 (2008); Phys. Rev. A. \textbf{81}, 063637 (2010).

\bibitem{Winkler} K. Winkler, G. Thalhammer, F. Lang, R. Grimm, J.
Hecker Denschlag, A.J. Daley, A. Kantian, H.P. B¡§uchler,
P. Zoller, Nature \textbf{441}, 853 (2006).

\end{thebibliography}
\end{document}